\documentclass[pra,preprint,superscriptaddress]{revtex4}

\usepackage{nicefrac,amsmath,amssymb}
\usepackage{graphicx}
\usepackage{bm}
\usepackage{xspace}
\usepackage{stmaryrd}
\usepackage{xifthen}

\newcommand{\bem}[0]{\textsc{bem}\xspace}

\newcommand{\mmatrix}[2][]{
  \ifthenelse{\isempty{#1}}
    {\left\llbracket{#2}\right\rrbracket}
    {#1\llbracket{#2}#1\rrbracket}
}
\newcommand{\dy}[1]{\overset\leftrightarrow{\bm #1}} 

\begin{document}

\title{Nanoscale electromagnetism with the boundary element method}

\author{Ulrich Hohenester}\thanks{E-mail \texttt{ulrich.hohenester@uni-graz.at}.}
\affiliation{Institute of Physics, University of Graz, Universit\"atsplatz 5, 8010 Graz, Austria}

\author{Gerhard Unger}
\affiliation{Institute of Applied Mathemathics, Technical University of Graz, Steyrergasse 30, 8010 Graz, Austria}

\date{February 8, 2022}

\begin{abstract}
In Yang et al. [Nature \textbf{576}, 248 (2019)], the authors introduced a general theoretical framework for nanoscale electromagnetism based on Feibelman parameters.  Here quantum effects of the optically excited electrons at the interface between two materials are lumped into two complex-valued and frequency-dependent parameters, which can be incorporated into modified boundary conditions for Maxwell's equations, the so-called mesoscopic boundary conditions.  These modifications can in principle be implemeted in any Maxwell solver, although the technicalities can be subtle and depend on the chosen computational approach.  In this paper we show how to implement the mesoscopic boundary conditions in a boundary element method approach, based on a Galerkin scheme with Raviart-Thomas shape elements for the representation of the tangential electromagnetic fields at the boundary.  We demonstrate that the results of our simulations are in perfect agreement with Mie theory including Feibelman parameters, and that for typical simulation scenarios the computational overhead is usually small.  
\end{abstract}
\maketitle

\section{Introduction}

Plasmonics has given photonics the ability to go to the nanoscale~\cite{barnes:03,atwater:07,schuller:10,novotny:11}.  This is achieved by optically exciting coherent electron charge oscillations at the boundary of metallic nanoparticles, so-called localized surface plasmon resonances or particle plasmons in short, which come along with strongly localized evanescent fields that allow focusing electromagnetic fields to deep subwavelength volumes~\cite{maier:07,hohenester:20}.  Applications are manifold and range from sensorics and photovoltaics, over catalysis and thermal management, to metamaterials.  By a similar token, hybrid photon-phonon excitations at the surface of ionic nanoparticles, so-called surface phonon polaritons, enable extreme light confinement in the infrared regime~\cite{caldwell:15}, which is of importance for the emerging fields of phononics~\cite{maldovan:13} and thermoelectrics~\cite{snyder:08}, as well as for the controlled heat transfer at the nanoscale~\cite{joulain:05,volokitin:07}.

The theoretical description of optical excitations of metallic, ionic, or dielectric nanoparticles is based on the solution of Maxwell's equations, where the optical response of the nanoparticles is usually modeled in terms of homogeneous, local, and isotropic permittivity and permeability functions.  The validity of such a classical description has been questioned from the early days of plasmonics~\cite{kreibig:95}, and it has become customary to coin the term ``quantum plasmonics'' for deviations from a purely classical description~\cite{tame:13,bozhevolnyi:17,mortensen:21}.  These include a nonlocal dielectric response~\cite{david:11,ciraci:12,luo:13,mortensen:14}, or quantum tunneling through sub-nanometer gaps between coupled nanoparticles~\cite{esteban:12,esteban:15}, which can lead to novel charge-transfer plasmons~\cite{savage:12}. 

Feibelman parameters provide a general and versatile scheme to account for modifications from a classical description at nanostructure interfaces~\cite{feibelman:82}.  They were first introduced by Feibelman in the description of reflection and transmission of plane waves at flat interfaces.  The basic idea is to model the wave propagation on both sides of the interface through solutions of Maxwell's equations using homogeneous and local permittivities, and to lump all quantum effects of the metal electrons in the vicinity of the surface, say in a region of about one nanometer, into two so-called Feibelman parameters.   These parameters $d_\perp(\omega)$, $d_\|(\omega)$ are usually complex valued and frequency dependent, and can be interpreted in terms of charge and current distribution displacements of the optically excited metal electrons~\cite{feibelman:82,hohenester:20,goncalves:20}.  Feibelman parameters were brought to the field of plasmonics in~\cite{teperik:13}, where the authors showed that they can accomodate $d$-band effects in transition metals and explain the somewhat counter-intuitive blue shift of surface plasmon resonances.

Recently, Yang et al.~\cite{yang:19} suggested a methodology to incorporate Feibelman parameters into a framework based on Maxwell's equations with modified boundary conditions
\begin{subequations}\label{eq:bcmeso}
\begin{eqnarray}
  D_2^\perp-D_1^\perp &=& d_\|\, \nabla_\|\cdot\left(\bm D_2^\|-\bm D_1^\|\right) \\
  B_2^\perp-B_1^\perp &=& 0 \\
  \bm E_2^\|-\bm E_1^\| &=& -d_\perp\nabla_\|\left(E_2^\perp-E_1^\perp\right)\\
  \bm H_2^\|-\bm H_1^\| &=& -i\omega d_\|\,\hat{\bm n}\times \left(\bm D_2^\|-\bm D_1^\|\right)\,,
\end{eqnarray}
\end{subequations}
where all modifications of the electronic response at the interface are encompassed in $d_\perp(\omega)$, $d_\|(\omega)$.  The beauty of Eq.~\eqref{eq:bcmeso} is that the quantum effects at the interface are described on the same footing as the bulk material properties, namely in terms of effective parameters, which are extracted from either experiment or first principles calculation.  While the bulk permittivity function is usually obtained from a coarse graining procedure~\cite{jackson:99}, which can incorporate quantum effects, the Feibelman parameters are obtained from quantum descriptions for the electron wavefunctions at the metal surface, using either simplified Drude or more realistic many-body models~\cite{feibelman:82,mortensen:21}.  The framework for the consideration of quantum effects within Maxwell's equations through Feibelman parameters has been denoted as ``nanoscale electromagnetism'', and the boundary conditions of Eq.~\eqref{eq:bcmeso} have been referred to as the ``mesoscopic boundary conditions''~\cite{yang:19}.

Although the solutions of Maxwell's equations with the mesoscopic boundary conditions have provided good agreement with experiment and complementary theoretical descriptions~\cite{yang:19,goncalves:20,mortensen:21}, the implementation of the mesoscopic boundary conditions into available home made or commerical Maxwell solvers proves to be non-trivial.  In~\cite{yang:19} the authors employed an iterative solution scheme using the standard boundary conditions of tangential electromagnetic fields~\cite{jackson:99}, where the right-hand sides of Eq.~\eqref{eq:bcmeso} were accounted for through additional surface charge and current distributions, which were successively updated until convergence was reached.  Alternatively, the auhors suggested a scheme based on resonance or quasinormal modes~\cite{leung:94,kristensen:12,sauvan:13,lalanne:19,kristensen:20}, where the modifications due to the mesoscopic boundary conditions were included in a kind of perturbation approach within lowest order.

In this paper, we develop a methodology for the solution of Maxwell's equations using a boundary element method (\bem) approach together with the mesoscopic boundary conditions of Eq.~\eqref{eq:bcmeso}, and implement the modifications in our home made Maxwell solver \textsc{nanobem}.  We demonstrate the applicability of our implementation for a few proof-of-principle simulations.  Quite generally, the boundary element method appears to be particularly well suited for the consideration of Feibelman parameters, as it precisely assumes solutions of the homogeneous Maxwell equations (with local and homogenous material properties) on both sides of an interface, and matches in a second step the fields across the interface using the boundary conditions.

We have organized our paper as follows.  In Sec.~\ref{sec:theory} we develop the methodology for nanoscale electromagnetism within the boundary element method approach.  We use the usual Stratton-Chu approach for the tangential electromagnetic fields~\cite{stratton:39,chew:95,hohenester:20}, together with a Galerkin scheme using Raviart-Thomas shape elements.  As in our implementation of the mesoscopic boundary conditions we will use some technicalities of the Galerkin scheme, we present the \bem methodology in more length than probably needed, mainly to keep our paper as self-contained as possible.  In Sec.~\ref{sec:results} we present a few selected examples for nanophotonics simulations including mesoscopic boundary conditions, and demonstrate that our results are in perfect agreement with Mie theory.  Finally, in Sec.~\ref{sec:summary} we discuss the computational cost of our approach and briefly summarize our work.

\section{Theory}\label{sec:theory}

\subsection{Boundary integral method}

The starting point of the boundary integral method is the dyadic Green's function

\begin{equation}\label{eq:gdyad}
  \dy G_j(\bm r,\bm r')=\left(\dy I+\frac{\nabla\nabla}{k_j^2}\right)
  \frac{e^{ik_j|\bm r-\bm r'|}}{4\pi|\bm r-\bm r'|}\,,
\end{equation}
where $k_j$ is a wavenumber, $\dy I$ the unit matrix, and we have used the outer product with $(\nabla\nabla)_{ij}=\partial_i\partial_j$.  Within an unbounded medium, the electric field $\bm E_{\rm inc}(\bm r)$ due to a current distribution $\bm J(\bm r')$ can then be expressed as~\cite{hohenester:20}
\begin{equation}\label{eq:einc}
  \bm E_{\rm inc}(\bm r)=i\mu_0\omega\int \dy G_j(\bm r,\bm r')\cdot\bm J(\bm r')\,d^3r'\,,
\end{equation}
where $\mu_0$ is the permeability of free space (we consider non-magnetic materials only), and $\omega$ is the angular frequency of the oscillating current distribution.  Thus, the Green's function is proportional to the electric field at position $\bm r$ generated by a unit current source at position $\bm r'$.  For reasons to become clear in a moment, we denote $\bm E_{\rm inc}(\bm r)$ as an ``incoming field''.  

\begin{figure}[t]
\centerline{\includegraphics[width=0.6\textwidth]{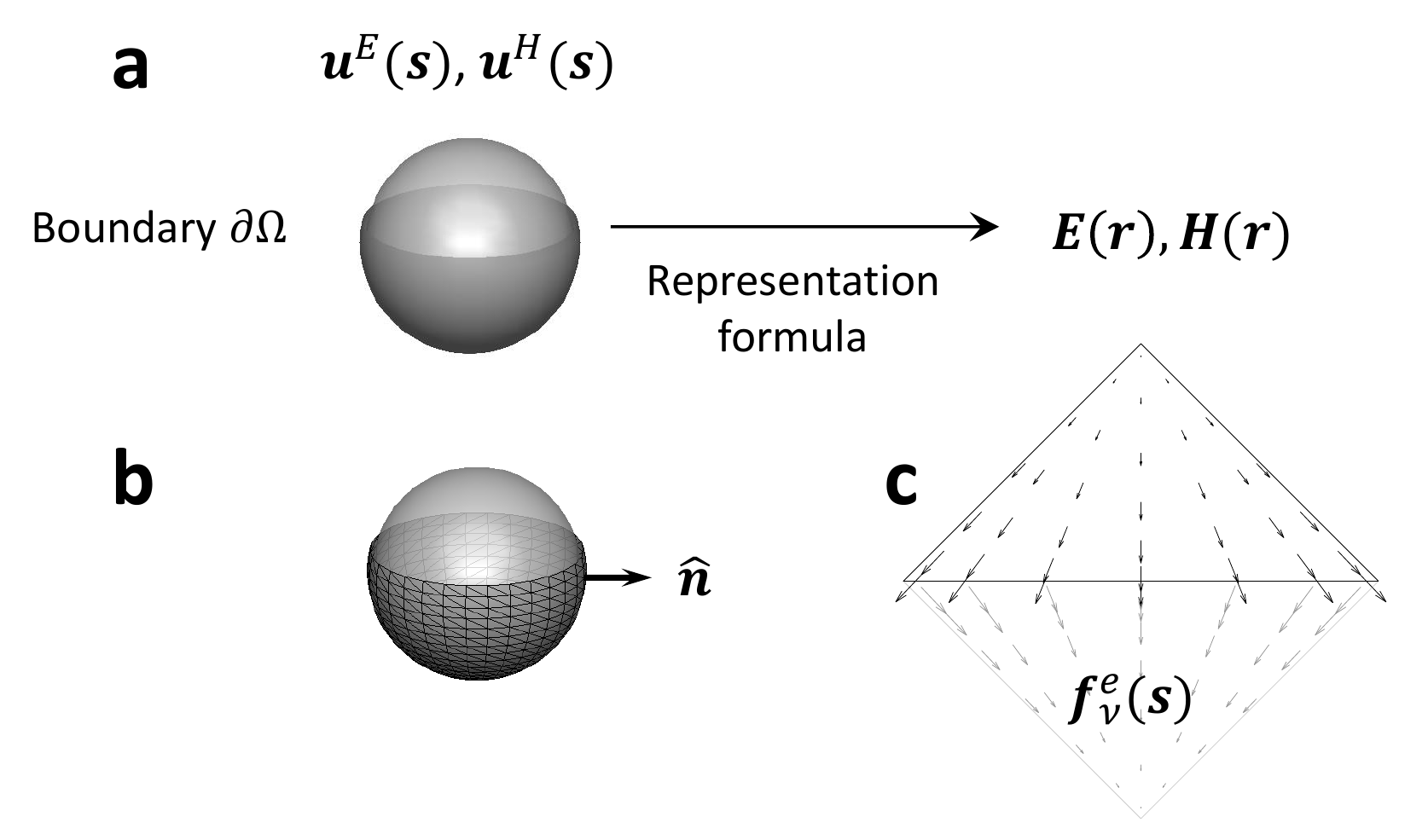}}
\caption{Schematics of (a) boundary integral method and (b) boundary element method.  (a) We consider a nanoparticle with a homogeneous and local permittivity function $\varepsilon_1(\omega)$ that is embedded in a background medium with permittivity $\varepsilon_2(\omega)$.  The sharp nanoparticle boundary is denoted with $\partial\Omega$ and the outer surface normal with $\hat{\bm n}$.  Once the tangential electromagnetic fields $\bm u^{E,H}$ are known at the boundary, they can be computed everywhere else using the representation formula of Eq.~\eqref{eq:rep}.  (b) In the boundary element method, the nanoparticle boundary is discretized using boundary elements of triangular shape, and (c) the tangential electromagnetic fields are approximated using Raviart-Thomas shape elements $\bm f_\nu^e$.  The working equations can be expressed in the form of matrix-vector multiplications, and the solutions $\bm u^{E,H}$ are obtained through matrix inversion.}\label{fig:schematics}
\end{figure}

In what follows, we consider the situation depicted in Fig.~\ref{fig:schematics} of a nanoparticle with permittivity $\varepsilon_1$, which may depend on frequency $\omega$, embedded in a background medium with permittivity $\varepsilon_2$.  In principle our formalism also applies to geometries of coupled or coated particles, as long as the permittivity functions $\varepsilon$ are local, homogeneous, and isotropic, but we here discuss the situation of a single particle only.  Following the seminal work of Stratton and Chu~\cite{stratton:39}, we express the electric field $\bm E(\bm r)$ outside the nanoparticle in terms of the tangential electromagnetic fields $\hat{\bm n}\times\bm E$, $\hat{\bm n}\times\bm H$ at the nanoparticle boundary $\partial\Omega$ through (see also \cite[Eq.~(5.26)]{hohenester:20})
\begin{equation}\label{eq:rep}
  \bm E(\bm r)=\bm E_{\rm inc}(\bm r)+\oint_{\partial\Omega}\left\{i\mu_0\omega\,\dy G_2(\bm r,\bm s')
  \cdot\hat{\bm n}'\times\bm H(\bm s')-
  \left[\nabla'\times\dy G_2(\bm r,\bm s')\right]\cdot\hat{\bm n}'\times\bm E(\bm s')\right\}\,dS'\,.
\end{equation}
Here $\bm E_{\rm inc}(\bm r)$ is the incoming field of Eq.~\eqref{eq:einc} produced by the current distribution in the embedding medium, $\hat{\bm n}$, $\hat{\bm n}'$ are the outer surface normals of the nanoparticle boundary at positions $\bm s$, $\bm s'$, and we denote positions on and off the boundary with $\bm s$ and $\bm r$, respectively.  Similar expressions can be obtained for the magnetic field and the electromagnetic fields inside the nanoparticle.  Eq.~\eqref{eq:rep} is reminiscent of Huygen's principle which propagates the fields at the wavefront (here $\hat{\bm n}\times\bm E$, $\hat{\bm n}\times\bm H$) to another position in space (here $\bm r$).  Thus, once $\hat{\bm n}\times\bm E$, $\hat{\bm n}\times\bm H$ are known at the boundary, we can compute the electromagnetic fields everywhere else using the representation formula of Eq.~\eqref{eq:rep}.

We next do some extra work and rewrite Eq.~\eqref{eq:rep} and the expressions for the remaining fields inside and outside the nanoparticle in a more compact form.  First, we introduce the abbreviations $\bm u_j^E(\bm s)=\hat{\bm n}\times\bm E(\bm s)$, $\bm u_j^H(\bm s)=\hat{\bm n}\times\bm H(\bm s)$ for the tangential electromagnetic fields at the boundary inside ($j=1$) and outside ($j=2$).  For the usual boundary conditions, the tangential electromagnetic fields are continuous when crossing the boundary, but for the modified boundary conditions of Eq.~\eqref{eq:bcmeso} the fields are discontinuous and we thus keep the index $j$ on $\bm u_j^{E,H}$.  We introduce the single and double layer potentials~\cite[Eq.~(5.34,35)]{hohenester:20}
\begin{subequations}\label{eq:layer}
\begin{eqnarray}
  \bigl[\mathbb{S}_j\bm u\bigr](\bm r)&=&\oint_{\partial\Omega}
  \dy G_j(\bm r,\bm s')\cdot\bm u(\bm s')\,dS'=\oint_{\partial\Omega}\left[
  g_j(\bm r,\bm s')\bm u(\bm s')+\frac 1{k_j^2}\nabla g_j(\bm r,\bm s')\,
  \nabla'\cdot\bm u(\bm s')\right]\,dS'\nonumber\\ &&\\
  \bigl[\mathbb{D}_j\bm u\bigr](\bm r)&=&\oint_{\partial\Omega}
  \nabla'\times\dy G_j(\bm r,\bm s')\cdot\bm u(\bm s')\,dS'=
  \oint_{\partial\Omega}\nabla'\times g_j(\bm r,\bm s')\bm u(\bm s')\,dS'\,,
\end{eqnarray}
\end{subequations}
where $g_j(\bm r,\bm r')$ is the scalar Green's function given by the fraction on the right-hand side of Eq.~\eqref{eq:gdyad}.  In obtaining the last expression in Eq.~(\ref{eq:layer}a) we have performed integration by parts and have used that the remaining contribution becomes zero for continuous tangential fields.  If we assume that the nanoparticle is excited only through sources located within the background medium, we obtain for the representation formula inside the nanoparticle~\cite{chew:95,hohenester:20}
\begin{subequations}\label{eq:rep2}
\begin{eqnarray}
  \bm E(\bm r) &=& -i\omega\mu_0\bigl[\mathbb{S}_1\bm u_1^H\bigr](\bm r)+
  \bigl[\mathbb{D}_1\bm u_1^E\bigr](\bm r)\\
  \bm H(\bm r) &=& +i\omega\varepsilon_1\bigl[\mathbb{S}_1\bm u_1^E\bigr](\bm r)+
  \bigl[\mathbb{D}_1\bm u_1^H\bigr](\bm r)\,.\qquad
\end{eqnarray}
Similarly, for positions $\bm r$ outside the particle the electromagnetic fields can be obtained from
\begin{eqnarray}
  \bm E(\bm r) &=& \bm E_2^{\rm inc}(\bm r)+i\omega\mu_0\bigl[\mathbb{S}_2\bm u_2^H\bigr](\bm r)-
  \bigl[\mathbb{D}_2\,\bm u_2^E\bigr](\bm r)\\
  \bm H(\bm r) &=& \bm H_2^{\rm inc}(\bm r)-i\omega\varepsilon_2\bigl[\mathbb{S}_2\bm u_2^E\bigr](\bm r)-
  \bigl[\mathbb{D}_2\bm u_2^H\bigr](\bm r)\,.\qquad
\end{eqnarray}
\end{subequations}
The reader might like to check that Eqs.~\eqref{eq:rep} and (\ref{eq:rep2}c) are indeed identical.  The representation formulas of Eq.~\eqref{eq:rep2} can be used for two purposes.  First, once the tangential electromagnetic fields are known at the boundary, we can compute the electromagnetic fields everywhere else.  Second, they can be used to determine the tangential fields at the boundary themselves.  To this end, we have to perform the limit $\bm r\to\bm s$ in Eq.~\eqref{eq:rep2}, where the position $\bm r$ approaches the boundary from either the inside or outside, and exploit the boundary conditions of the fields.  While the limit can be performed safely for the single layer potential, $\lim_{\bm r\to\bm s}\hat{\bm n}\times\bigl[\mathbb{S}_j\bm u\bigr](\bm r)=\hat{\bm n}\times\bigl[\mathbb{S}_j\bm u\bigr](\bm s)$, in the evaluation of the double layer potential we have to be careful on whether we approach the boundary from the inside or outside~\cite{hohenester:20}
\begin{equation}
  \lim_{\bm r\to\bm s}\hat{\bm n}\times\bigl[\mathbb{D}_{1,2}\bm u\bigr](\bm r)=
  \pm\mbox{$\frac 12$}\bm u(\bm s)+\hat{\bm n}\times\bigl[\mathbb{D}_{1,2}\bm u\bigr](\bm s)\,.
\end{equation}
Here the positive sign has to be taken for the limit from the inside, and the negative sign for the limit from the outside.  Thus, if we consider in Eqs.~(\ref{eq:rep2}a,b) the tangential fields $\hat{\bm n}\times\bm E$, $\hat{\bm n}\times\bm H$ and approach the boundary from the inside, we get
\begin{subequations}\label{eq:rep3}
\begin{eqnarray}
  \mbox{$\frac 12$}\bm u_1^E(\bm s) &=& -i\omega\mu_0\hat{\bm n}\times\bigl[\mathbb{S}_1\bm u_1^H\bigr](\bm s)+
  \hat{\bm n}\times\bigl[\mathbb{D}_1\bm u_1^E\bigr](\bm s)\\
  \mbox{$\frac 12$}\bm u_1^H(\bm s) &=& +i\omega\varepsilon_1\hat{\bm n}\times\bigl[\mathbb{S}_1\bm u_1^E\bigr](\bm s)+
  \hat{\bm n}\times\bigl[\mathbb{D}_1\bm u_1^H\bigr](\bm s)\,.\qquad
\end{eqnarray}
Similarly, we obtain from Eqs.~(\ref{eq:rep2}c,d)
\begin{eqnarray}
  \mbox{$\frac 12$}\bm u_2^E(\bm s) &=& \hat{\bm n}\times\bm E_2^{\rm inc}(\bm s)+
  i\omega\mu_0\hat{\bm n}\times\bigl[\mathbb{S}_2\bm u_2^H\bigr](\bm s)-
  \hat{\bm n}\times\bigl[\mathbb{D}_2\bm u_2^E\bigr](\bm s)\\
  \mbox{$\frac 12$}\bm u_2^H(\bm s) &=& \hat{\bm n}\times\bm H_2^{\rm inc}(\bm s)-
  i\omega\varepsilon_2\hat{\bm n}\times\bigl[\mathbb{S}_2\bm u_2^E\bigr](\bm s)-
  \hat{\bm n}\times\bigl[\mathbb{D}_2\bm u_2^H\bigr](\bm s)\,.\qquad
\end{eqnarray}
\end{subequations} 
Below we will show how to transform Eqs.~\eqref{eq:rep3} into matrix equations using a boundary element method approach.  Before doing so, we rewrite the boundary conditions of Eq.~\eqref{eq:bcmeso} using the tangential electromagnetic fields. The identity~\cite[Cor.~3.16]{assous:18}
\begin{equation*}
\hat{\bm n}  \cdot\nabla \times\bm u
=-\nabla_\|\cdot(\hat{\bm n}\times\bm u)
\end{equation*}
and the curl equations $\nabla\times\bm E=i\mu_0\omega\bm H$, $\nabla\times\bm H=-i\varepsilon\omega\bm E$ enable us to express the normal components $E^\perp$, $H^\perp$  as
\begin{subequations}
\begin{eqnarray}
  E_j^\perp=\hat{\bm n}\cdot\bm E_j&=&-\frac i{\varepsilon_j\omega}\nabla_\|\cdot\bm u_j^H\\
  H_j^\perp=\hat{\bm n}\cdot\bm H_j&=&\phantom-\frac i{\mu_0\omega}\nabla_\|\cdot\bm u_j^E\,.
\end{eqnarray}
\end{subequations}
With this, the mesoscopic boundary conditions of Eq.~(\ref{eq:bcmeso}c,d) for the tangential electromagnetic fields can be cast to the form
\begin{subequations}\label{eq:bcmeso2}
\begin{eqnarray}
  \bm u^E_2-\bm u^E_1 &=& \frac{id_\perp}\omega \hat{\bm n}\times\nabla_\|\nabla_\|\cdot\left(
  \frac{\bm u^H_2}{\varepsilon_2}-\frac{\bm u^H_1}{\varepsilon_1}\right)\\
  \bm u^H_2-\bm u^H_1 &=& -i\omega d_\|\,\hat{\bm n}\times 
  \left(\varepsilon_2\bm u^E_2-\varepsilon_1\bm u^E_1\right)\,.
\end{eqnarray}
\end{subequations}
The remaining boundary conditions for the normal components are obtained by taking on both sides of Eq.~\eqref{eq:bcmeso2} the divergence $\nabla_\|\cdot$ along the boundary directions.

\subsection{Boundary element method}

We next submit the boundary integral equations~\eqref{eq:rep3} and the boundary conditions of Eq.~\eqref{eq:bcmeso2} to a boundary element method (\bem) and a Galerkin scheme.  For that, let us introduce the pairing
\begin{equation}\label{eq:pairing}
  \big<\bm w,\bm u\big>=\oint_{\partial\Omega} \bm w(\bm s)\cdot\bm u(\bm s)\,dS\,,
\end{equation}
where $\bm w(\bm s)$ is an arbitrary tangential vector function.  We use this pairing  to bring the boundary integral equations~\eqref{eq:rep3} and the boundary conditions~\eqref{eq:bcmeso2} to a variational form which is necessary in order to apply a Galerkin method.
Eq.~(\ref{eq:rep3}a) is considered in the form
\begin{subequations}\label{eq:reppair}
\begin{eqnarray}
0 
&=& 
\left<\hat{\bm n}\times\bm w,
  \mbox{$\frac 12$}\bm u_1^E+i\omega\mu_0\hat{\bm n}\times\mathbb{S}_1\bm u_1^H-\hat{\bm n}\times\mathbb{D}_1\bm u_1^E\right> = \nonumber\\ &=&
\left<\bm w,
  -\mbox{$\frac 12$}\hat{\bm n}\times\bm u_1^E+i\omega\mu_0\mathbb{S}_1\bm u_1^H-\mathbb{D}_1\bm u_1^E\right>
  \,,
\end{eqnarray}
where we have used $\left<\bm w,\hat{\bm n}\times\bm f\right>=-\left<\hat{\bm n}\times\bm w,\bm f\right>$ and the identity $\hat{\bm n}\times\hat{\bm n}\times\bm w=-\bm w$. Similarly we get for Eqs.~(\ref{eq:rep3}b--d)
\begin{eqnarray}
\left<\bm w,
  -\mbox{$\frac 12$}\hat{\bm n}\times\bm u_1^H
  -i\omega\varepsilon_1\mathbb{S}_1\bm u_1^E-\mathbb{D}_1\bm u_1^H\right>&=&0\,,
\\
\left<\bm w,
  -\mbox{$\frac 12$}\hat{\bm n}\times\bm u_2^E-i\omega\mu_0\mathbb{S}_2\bm u_2^H+\mathbb{D}_2\bm u_2^E\right>
  &=& \left<\bm w, \bm E_2^{\rm{inc}}  \right>
\\
\left<\bm w,
  -\mbox{$\frac 12$}\hat{\bm n}\times\bm u_2^H 
  +i\omega\varepsilon_2\mathbb{S}_2\bm u_2^E+\mathbb{D}_2\bm u_2^H\right>
  &=& \left<\bm w, \bm H_2^{\rm{inc}}  \right>\,.
 \end{eqnarray}
\end{subequations}
Using the pairing of Eq.~\eqref{eq:pairing}, we can employ integration by parts~\cite[Sect.~3.1]{assous:18} in order shuffle around derivatives via
\begin{equation}\label{eq:shuffle}
  \oint_{\partial\Omega} \bm w(\bm s)\cdot\nabla_\| u(\bm s)\,dS=
  -\oint_{\partial\Omega}\big(\nabla_\|\cdot\bm w(\bm s)\big) u(\bm s) dS=
  -\big<\nabla_\|\cdot\bm w,u\big>\,,
\end{equation}
where $u$ is some scalar function, and in the last equality we have introduced a shorthand notation.  This can be used to represent the  single layer potential of Eq.~(\ref{eq:layer}a) on the boundary as
\begin{equation}
  \big<\bm w,\mathbb{S}_j\bm u\big>=\oint_{\partial\Omega}\left[
  \bm w(\bm s)\cdot \bm u(\bm s')-\frac 1{k_j^2}
  \Big(\nabla_\|\cdot\bm w(\bm s)\Big)\Big(\nabla_\|\cdot\bm u(\bm s')\Big)
  \right]g_j(\bm s,\bm s')\,dSdS'\,.
\end{equation}
Submitting the boundary conditions of Eq.~\eqref{eq:bcmeso2} to the pairing leads us to
\begin{eqnarray}\label{eq:bctest}
  \big<\bm w,\bm u^E_1\big>-\frac{id_\perp}{\omega\varepsilon_1} 
  \big<\nabla_\|\cdot\hat{\bm n}\times\bm w,\nabla_\|\cdot\bm u^H_1\big> &=&
  \big<\bm w,\bm u^E_2\big>-\frac{id_\perp}{\omega\varepsilon_2}
  \big<\nabla_\|\cdot\hat{\bm n}\times\bm w,\nabla_\|\cdot\bm u^H_2\big>
  \nonumber\\
  \big<\bm w,\bm u^H_1\big>+i\omega\varepsilon_1 d_\|\,
\big<\bm w,\hat{\bm n}\times\bm u^E_1\big> &=&  \big<\bm w,\bm u^H_2\big>+i\omega\varepsilon_2 d_\|\,\big<\bm w,\hat{\bm n}\times\bm u^E_2\big>\,,
\end{eqnarray}
where we have again employed Eq.~\eqref{eq:shuffle} to simplify the $\nabla_\|\nabla_\|\cdot$ term, which would be hard to handle in a computational approach otherwise.  In the \bem approach, we approximate the boundary through a discretization in terms of triangular boundary elements $\tau_i$,
\begin{equation}
  \partial\Omega\approx\bigcup_i\tau_i\,.
\end{equation}
For simplicity we assume that all boundary elements $\tau_i$ have a triangular shape, although our approach would also work for other discretizations, e.g., using quadrilateral or mixed shapes.  We additionally discretize the tangential fields $\bm u$ through Raviart-Thomas or Rao-Wilton-Glisson basis elements, see Fig.~\ref{fig:schematics}, which guarantee continuity of the tangential fields when going from one triangle to an adjacent one~\cite{chew:95,hohenester:20}.  Technically, this is done by assigning to each edge $\nu$ of the discretized boundary a value for $\mmatrix[\normalsize]u_\nu$ and by using tangential basis functions $\bm f_\nu^e$ that are nonzero in the two adjacent triangles only (a so-called local support), and which are constructed such that the outflow from one triangle equals the inflow to the other triangle.  For details see~\cite{chew:95,hohenester:20}.  The tangential fields can then be approximated through
\begin{equation}\label{eq:shape}
  \bar{\bm u}^{E,H}_j(\bm s)=\sum_{\nu=1}^n \bm f_\nu^e(\bm s)\,\mmatrix[\big]{u_j^{E,H}}_\nu\,,
\end{equation}
where $n$ is the total number of individual edges that determines the number of degrees of freedom for the \bem approach.  Within the Galerkin scheme, we insert the functions $\bar{\bm u}^{E,H}_j$ instead of $\bm u^{E,H}_j$ in the Eqs.~\eqref{eq:reppair} and~\eqref{eq:bctest}, and use as test functions the basis functions $\bm f_\nu^e$. This  gives 
\begin{subequations}\label{eq:GalerkinReppair}
 \begin{eqnarray}
\left<\bm f_\nu^e,
  -\mbox{$\frac 12$}\hat{\bm n}\times\bar{\bm u}_1^E+i\omega\mu_0\mathbb{S}_1\bar{\bm u}_1^H-\mathbb{D}_1\bar{\bm u}_1^E\right>
&=& 0  
\\
\left<\bm f_\nu^e,
  -\mbox{$\frac 12$}\hat{\bm n}\times\bar{\bm u}_1^H
  -i\omega\varepsilon_1\mathbb{S}_1\bar{\bm u}_1^E-\mathbb{D}_1\bar{\bm u}_1^H\right>
 &=& 0  
\\
\left<\bm f_\nu^e,
  -\mbox{$\frac 12$}\hat{\bm n}\times\bar{\bm u}_2^E-i\omega\mu_0\mathbb{S}_2\bar{\bm u}_2^h+\mathbb{D}_2\bar{\bm u}_2^E\right>
&=&  \left<\bm f_\nu^e, \bm E_2^{\rm{inc}}  \right>
\\
\left<\bm f_\nu^e,
  -\mbox{$\frac 12$}\hat{\bm n}\times\bar{\bm u}_2^H 
  +i\omega\varepsilon_2\mathbb{S}_2\bar{\bm u}_2^E+\mathbb{D}_2\bar{\bm u}_2^H\right>
&=& 
\left<\bm f_\nu^e, \bm H_2^{\rm{inc}}  \right>\,,
\end{eqnarray}
\end{subequations}
and
 \begin{eqnarray}\label{eq:Galerkinbctest}
 \big<\bm f_\nu^e,\bar{\bm u}^E_1\big>-\frac{id_\perp}{\omega\varepsilon_1} 
  \big<\nabla_\|\cdot\hat{\bm n}\times\bm f_\nu^e,\nabla_\|\cdot\bar{\bm u}^H_1\big> &=&
 \big<\bm f_\nu^e,\bar{\bm u}^E_2\big>-\frac{id_\perp}{\omega\varepsilon_2} 
  \big<\nabla_\|\cdot\hat{\bm n}\times\bm f_\nu^e,\nabla_\|\cdot\bar{\bm u}^H_2\big>
  \nonumber\\
  \big<\bm f_\nu^e,\bar{\bm u}^H_1\big>
  +i\omega\varepsilon_1 d_\|\,
  \big<\bm f_\nu^e,
  \hat{\bm n}\times \bar{\bm u}^E_1\big> &=& 
  \big<\bm f_\nu^e,\bar{\bm u}^H_2\big>+i\omega\varepsilon_2 d_\|\,\big<\bm f_\nu^e,\hat{\bm n}\times\bar{\bm u}^E_2\big>\,.
\end{eqnarray}
These equations are used to determine the unknown   expansion coefficients $\mmatrix[\normalsize]{u^{E,H}}_\nu$.
Combining the electric and magnetic components of the solution vectors according to
\begin{equation}\label{eq:uvec}
  u_j=\begin{pmatrix} \mmatrix[\normalsize]{u_j^E} \\ \mmatrix[\normalsize]{u_j^H} \end{pmatrix}\,,
\end{equation}
we rewrite Eq.~\eqref{eq:GalerkinReppair} in the compact form
\begin{subequations}\label{eq:cal}
\begin{eqnarray}
  \left(\frac 12I-A_1\right)u_1 &=& 0 \\
  \left(\frac 12I+A_2\right)u_2 &=& q.
\end{eqnarray}
\end{subequations}
A more detailed discussion, as well as a definition of the  matrices $I$ and   $A_j$ is given in Appendix~\ref{app:working}.  Eq.~(\ref{eq:cal}) is usually referred to as the Calderon identities.  In order to solve for the unknowns $u_1$, $u_2$, we have to combine the two equations and invoke the boundary conditions of Eq.~\eqref{eq:Galerkinbctest}, which can be cast to the form
\begin{equation}\label{eq:bcworking}
  B_1u_1=B_2u_2\,,
\end{equation}
with the matrices $B_1$, $B_2$ given in Eq.~\eqref{eq:matB}.  This expression together with the Calderon identities of Eq.~\eqref{eq:cal} allow us to obtain the desired solutions.  We here introduce a scheme that is inspired by the Poggio-Miller-Chang-Harrington-Wu-Tsai formulation~\cite{chang:77,poggio:73,wu:77} where the two Calderon identities are subtracted
\begin{equation}\label{eq:working}
  \left[\left(\frac 12I+A_2\right)-\left(\frac 12I-A_1\right)B_1^{-1}B_2^{\phantom{1}}\right]u_2=q\,.
\end{equation}
The $B_1^{-1}B_2^{\phantom{1}}$ term accounts for the modified boundary conditions, which would become one for the case where both Feibelman parameters are set to zero.  Thus, the solution of the \bem equations, namely the inversion of the term in brackets of Eq.~\eqref{eq:working}, is highly similar to the usual solution scheme.  Once the tangential electromagnetic fields $u_2$ are known, we can compute the fields everywhere in the embedding medium using the representation formulas of Eq.~\eqref{eq:rep2}.  The solution $u_1$ inside the nanoparticle can be obtained through $u_1=B_1^{-1}B_2^{\phantom{1}}u_2$.

\section{Results}\label{sec:results}

\begin{figure}[t]
\includegraphics[width=\textwidth]{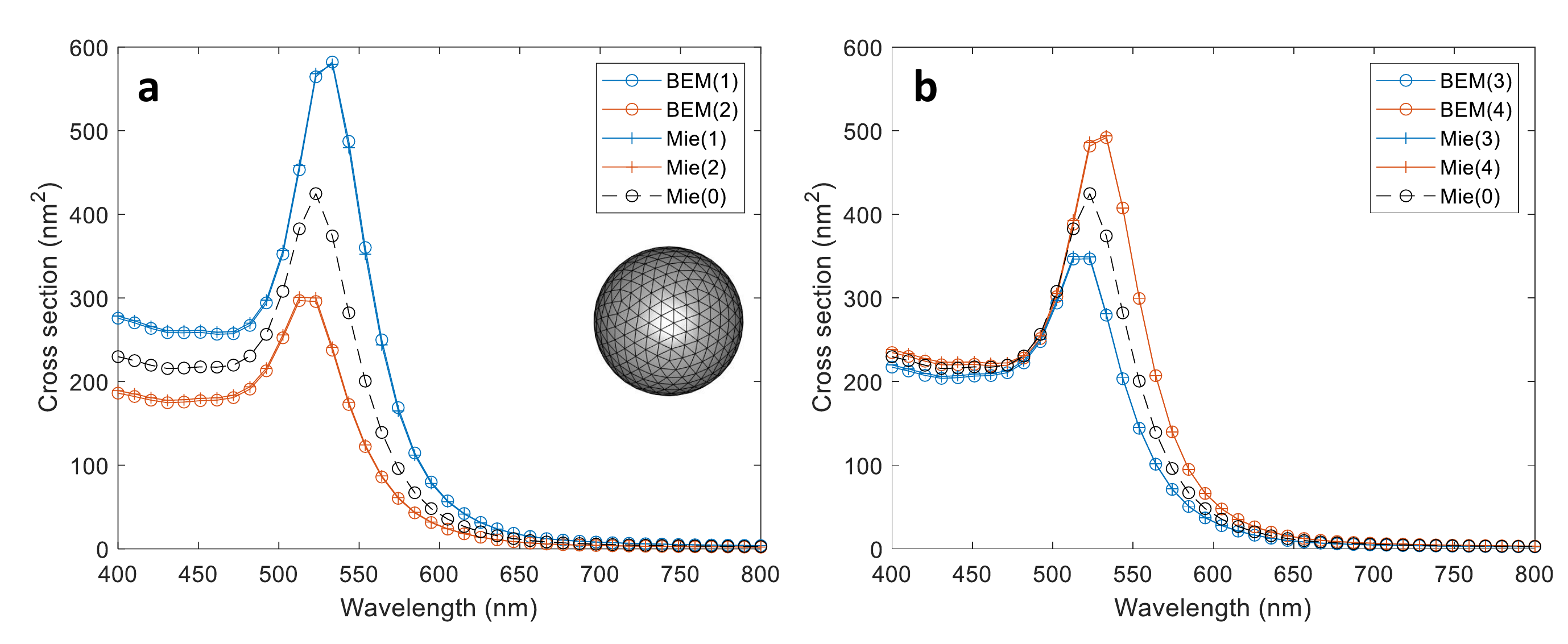}
\caption{Extinction cross sections for gold nanosphere (diameter 20 nm, dielectric function taken from~\cite{johnson:72}) embedded in water (refractive index $n_b=1.33$).  We use large and frequency independent Feibelman parameters of (1) $d_\perp=0.5$ nm, $d_\|=0$, (2) $d_\perp=-0.5$ nm, $d_\|=0$, (3) $d_\perp=0$, $d_\|=0.5$ nm, (4) $d_\perp=0$, $d_\|=-0.5$ nm.  The circle symbols report results of our \bem approach using the sphere discretization with 400 vertices shown in the inset of panel (a), the cross symbols report results of Mie theory including Feibelman parameters.  \bem and Mie results are in perfect agreement throughout.  The dashed line shows for comparison results of standard Mie theory without Feibelman parameters.  
}
\label{fig:sphere}
\end{figure}

We have implemented the working equation~\eqref{eq:working} in our home made \bem solver \textsc{nanobem}~\cite{hohenester:21}.  Fig.~\ref{fig:sphere} shows results for an optically excited gold nanosphere with 20 nm diameter, which is embedded in water.  For the permittivity function we use tabulated values extracted from optical experiment~\cite{johnson:72}.  We set the Feibelman parameters $d_\perp$, $d_\|$ to constant but otherwise arbitrary values, which are reported in the figure caption, and compare our results with those of a Mie theory including Feibelman parameter~\cite{goncalves:20}.  As can be seen in the figure, the results of our \bem simulations and Mie theory are in perfect agreement and almost indistinguishable throughout, thus demonstrating the accuracy of our  computational approach.  We will comment on the performance of our modified \bem approach further below in Sec.~\ref{sec:summary}.  Quite generally, with this proof-of-principle results we are now in the position to perform simulations including Feibelman parameters for any other setup that can be modeled within a \bem approach.  In the following we discuss two simple setups, namely coupling of nanoparticles and the computation of resonance modes, mainly to demonstrate the potential of our scheme.  More detailed investigations, including also Feibelman parameters extracted from ab-initio calculations, will be presented elsewhere.

\begin{figure}[t]
\centerline{\includegraphics[width=0.7\textwidth]{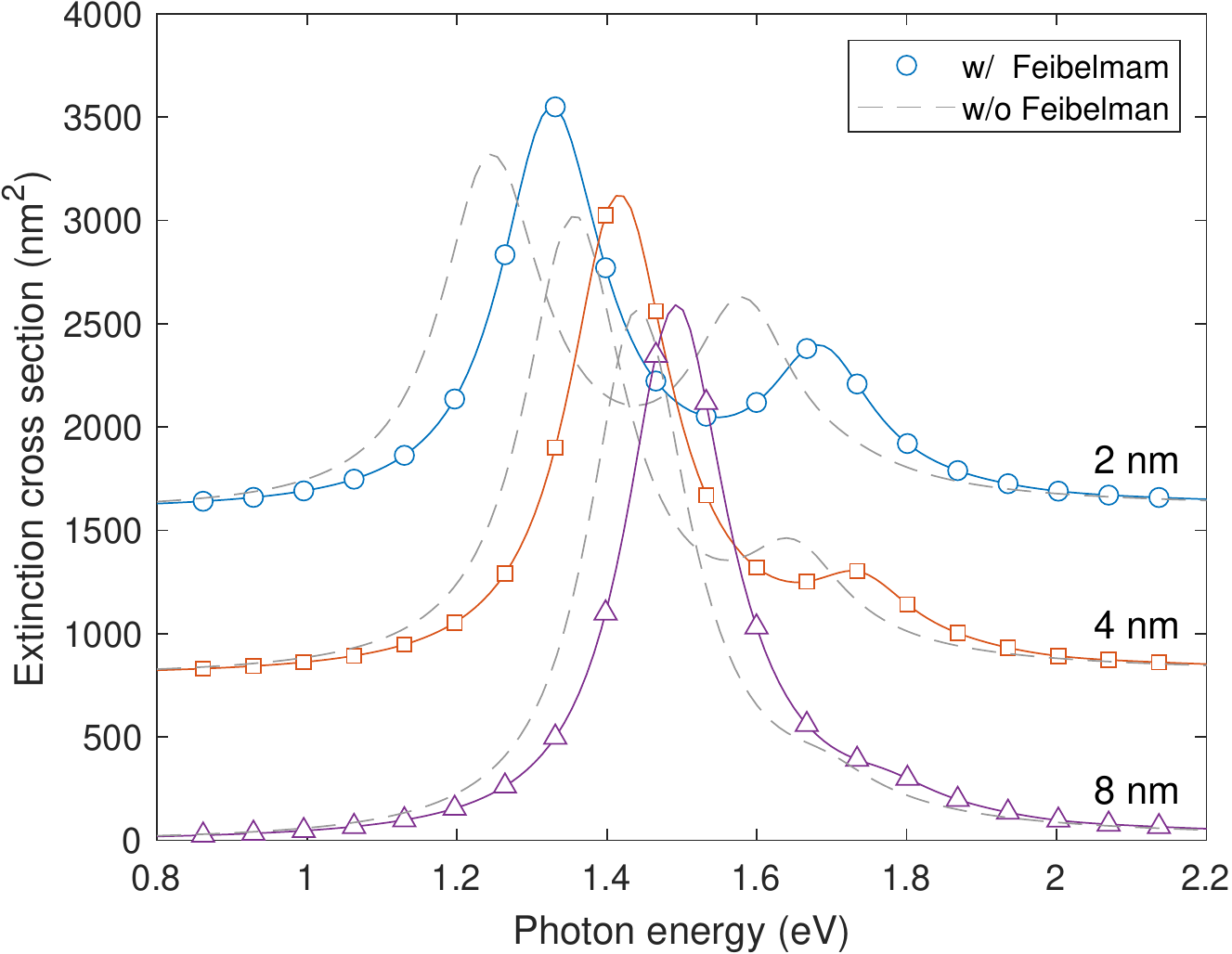}}
\caption{Extinction cross section for coupled gold nanospheres with a diameter of 20 nm and for gap distances of 2, 4, and 8 nm, as indicated on the right-hand side (we use $n_b=1.33$).  The spectra are offset for clarity, and the polarization of the incoming light is along the symmetry axis of the coupled spheres.  The two peaks are associated with bonding and anti-bonding dimer modes, and the splitting increases with decreasing gap distance owing to the increased coupling strength.}\label{fig:coupled}
\end{figure}

In the following we use a Drude dielectric function representative for gold~\cite{luo:13}, with a plasma frequency of $\hbar\omega_p=3.3$ eV and a damping constant $\hbar\gamma=0.165$ eV, together with the Feibelman parameters for the hydrodynamic model~\cite{feibelman:82}
\begin{equation}
  d_\perp(\omega)=-\frac{\beta}{\sqrt{\omega_p^2-\omega_{\phantom p}^2}}\,,\quad d_\|=0\,.
\end{equation}
The $\beta$ parameter accounts for the hydrodynamic pressure of an electron gas, its value of $\beta=0.0063\,c$ is  taken from Ref.~\cite{luo:13}.  Fig.~\ref{fig:coupled} shows the optical spectra for two coupled gold nanospheres and for different gap distances.  One observes two peaks associated with the bonding and antibonding dimer modes~\cite{hohenester:20}, and the peak splitting increases with decreasing gap distance owing to the enhanced coupling between the spheres.  For the chosen parameters, the simulation results with (solid lines) and without (dashed lines) consideration of Feibelman parameters are similar, apart from an approximately constant shift.

\begin{figure}[t]
\centerline{\includegraphics[width=0.7\textwidth]{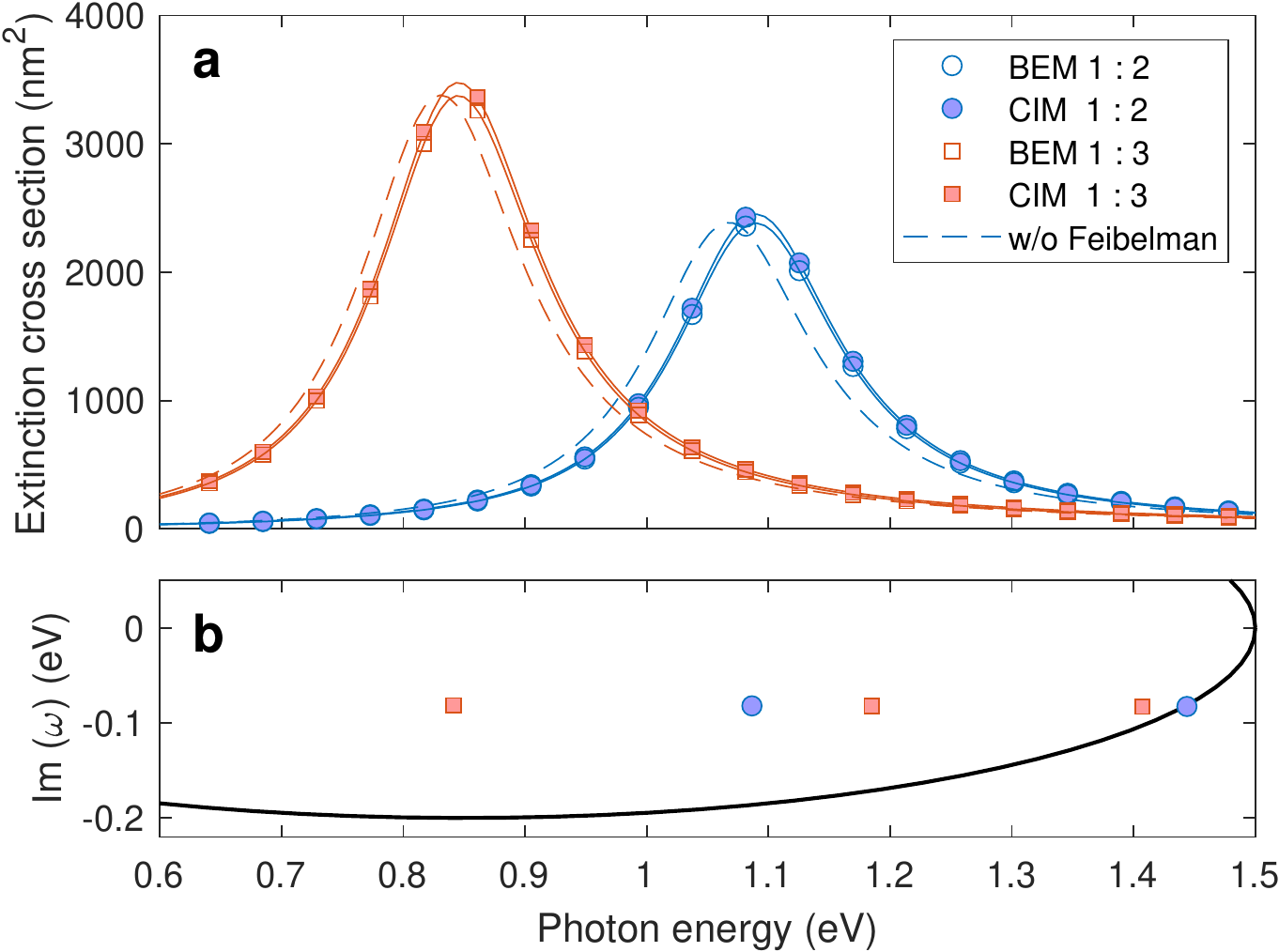}}
\caption{Computation of resonance modes and optical spectra for nanoellipsoids using a dielectric function representative for gold~\cite{luo:13}.  (a) Extinction cross sections for prolate nanoellipsoids with a short axis of 20 nm and a long axis of 40 nm (circles), and 60 nm (squares), the polarization of the incoming light is parallel to the long axis.  The open symbols show results from the full \bem simulations, the open symbols from the resonance mode approximation, and the dashed lines from simulations without Feibelman parameters.  (b) Complex frequency plane.  The solid line shows the contour used in the computation of the resonance modes, and the symbols the location of the resonance mode energies.}\label{fig:cim}
\end{figure}

As a final expample, in Fig.~\ref{fig:cim} we show results of simulations with resonance or quasinormal modes~\cite{leung:94,kristensen:12,sauvan:13,lalanne:19,kristensen:20} for gold nanoellipsoids, following the prescription given in~\cite{unger:18,hohenester:21}.  Importantly, the calculation of the resonance modes is almost identical for simulations with and without mesoscopic boundary conditions, with the exception of the additional boundary matrices to be considered in Eq.~\eqref{eq:working}~\cite{comment.cim}.  Panel (b) shows the complex resonance energies obtained from our contour integral method~\cite{unger:18,hohenester:21}, and panel (a) the extinction spectra obtained from the full \bem simulations (open symbols) and the resonance mode expansionss (full symbols).  We observe that the spectra are in perfect agreement.  The simulation results presented in Figs.~\ref{fig:coupled} and \ref{fig:cim} demonstrate that all simulations that can be performed with standard \bem solvers can indeed be equally well performed with \bem solvers incorporating mesosocopic boundary conditions.

\section{Discussion and Summary}\label{sec:summary}

We finally comment on the computer times for \bem simulations with and without mesoscopic boundary conditions, which is related to the additional computation of the $B_1^{-1}B_2^{\phantom 1}$ term in Eq.~\eqref{eq:working}.  For coarse boundary discretizations with a few hundred boundary elements, the main computational cost is the evaluation of the single and double layer potentials, see Eq.~\eqref{eq:layer2}, and the evaluation and inversion of the boundary matrices $B_1$, $B_2$ leads to no significant overhead.  Things may change for finer discretizations with a few thousand boundary elements, where simulations can be slowed down by a factor between two and three in comparison to normal \bem simulations.  For nanoparticles with separated boundaries, such as for coupled or coated particles, the evaluation of $B_1^{-1}B_2^{\phantom 1}$ can be done blockwise, which leads to a significant speedup and comparable computer times for simulations with and without mesoscopic boundary conditions.  Thus, the additional overhead in Eq.~\eqref{eq:working} is usually small.


\begin{figure}[t]
\centerline{\includegraphics[width=0.7\textwidth]{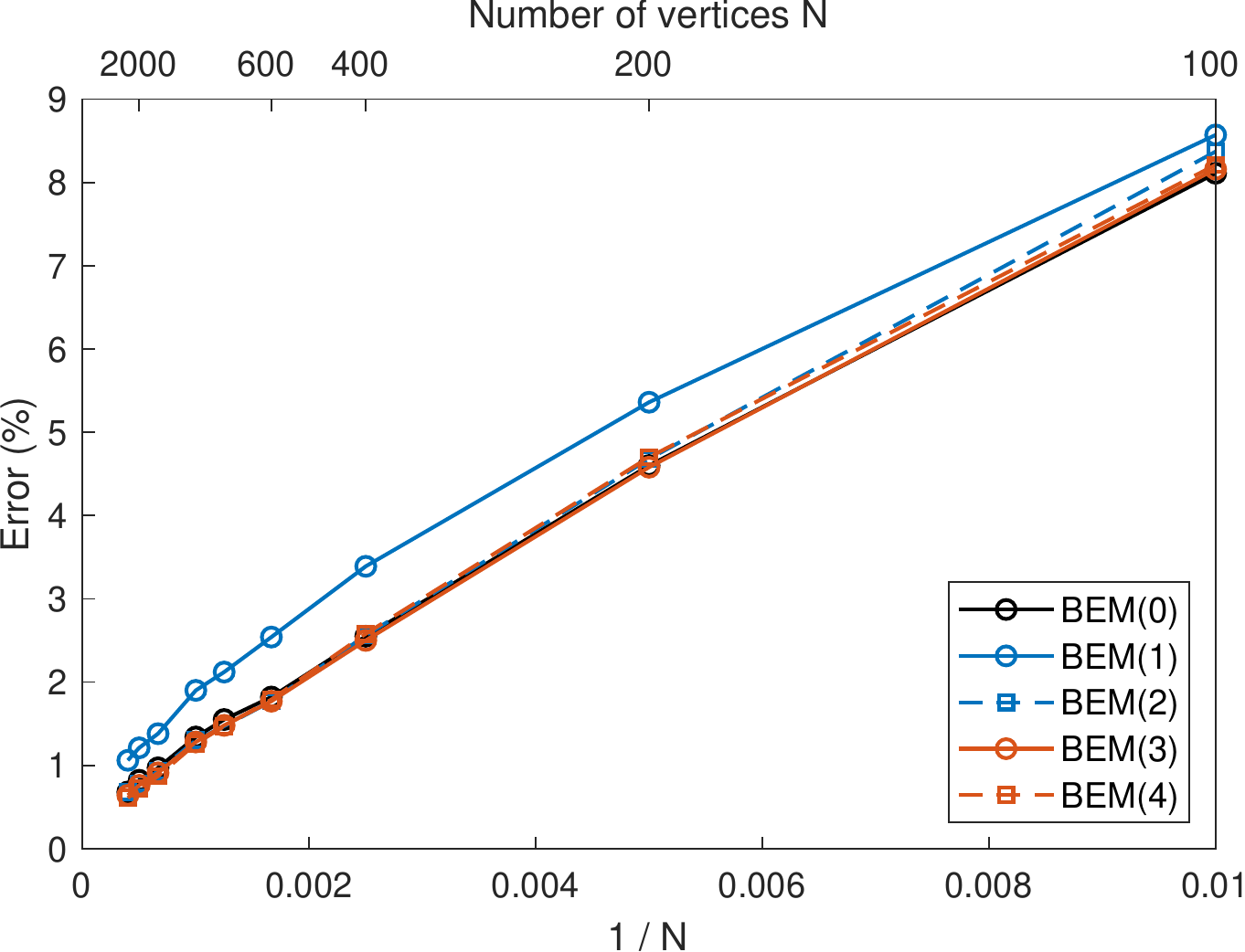}}
\caption{Error between tangential electric fields computed within Mie theory and \bem simulations, evaluated at the centroids of the boundary elements, see Eq.~\eqref{eq:error}.  We consider a gold nanosphere with 20 nm diameter and the same material and simulation parameters as listed in the caption of Fig.~\ref{fig:sphere}, and compare simulation results for boundary discretizations with a varying number of vertices $N$ and for a wavelength of 520 nm.  With decreasing mesh size the error becomes smaller monotonously.}\label{fig:error}
\end{figure}

In Fig.~\ref{fig:error} we investigate the accuracy of our \bem implementation.  We consider a gold nanosphere with 20 nm diameter and use the same material and simulation parameters as listed in the caption of Fig.~\ref{fig:sphere}, and compare simulation results for boundary discretizations with a varying number of vertices $N$ and correspondingly boundary elements.  For each discretization we compute the tangential electric fields at the centroids of the boundary elements and evaluate the deviations from the exact fields obtained within Mie theory,
\begin{equation}\label{eq:error}
  \mbox{error}=\sqrt{\frac{
  \sum_{i}\big|\hat{\bm n}_i\times\left(\bm E_i^{\rm BEM}-\bm E_i^{\rm Mie}\right)\big|^2}
  {\sum_{i}^n\big|\hat{\bm n}_i\times\bm E_i^{\rm Mie}\big|^2}}
\end{equation}
As can be seen in Fig.~\ref{fig:error}, the error decreases monotonously when increasing the number of vertices, where all simulation setups exhibit a similar error slope.  This demonstrates the accuracy and robustness of the scheme described in this work.

To summarize, we have presented a methodology for the implementation of mesoscopic boundary conditions within a \bem approach, and have demonstrated that the results of such an approach are in perfect agreement with Mie theory.  Further case studies have revealed that our implementation can be used in all situations where normal \bem simulations can be employed, and that the computational overhead is negligible to moderate in most cases of interest.  This eshablishes \bem as a viable and efficient solution scheme for nanoscale electromagnetism including mesoscopic boundary conditions.

\section*{Acknowledgements}

We thank Asger Mortensen for helpful discussions and for suggesting the implementation of the mesoscopic boundary conditions within a boundary element method approach.  This work has been supported in part by the Austrian Science Fund FWF under project P 31264 and by NAWI Graz.

\begin{appendix}

\section{}\label{app:working}

In this Appendix we provide further details for the derivation of Eqs.~\eqref{eq:cal} and \eqref{eq:bcworking}.  With the shape functions of Eq.~\eqref{eq:shape}, we introduce the matrix elements for the discretized single and double layer potentials
\begin{subequations}\label{eq:layer2}
\begin{eqnarray}
  \mmatrix[\big]{\mathbb S}_{\nu\nu'}
  &=& \oint\oint \left[
  \bm f_{\nu}^e(\bm s)\cdot\bm f_{\nu'}^e(\bm s')-
  \frac{\nabla_\|\cdot\bm f_{\nu}^e(\bm s)\,\nabla_\|'\cdot\bm f_{\nu'}^e(\bm s')}{k^2}\right]g(\bm s,\bm s')\,dSdS'\qquad\\
  \mmatrix[\big]{\mathbb D}_{\nu\nu'}&=&\oint\oint
  \bm f_{\nu}^e(\bm s)\cdot\nabla' g(\bm s,\bm s')\times\bm f_{\nu'}^e(\bm s')\,dSdS'\,,
\end{eqnarray}
\end{subequations}
together with the  matrix
\begin{equation}\label{eq:matI}
  \mmatrix[\big]{I}_{\nu\nu'}=\big<\bm f_\nu^e,\hat{\bm n}\times\bm f_{\nu'}^e\big>\,.
\end{equation}
For the incoming fields, we introduce the inhomogeneities
\begin{subequations}
\begin{eqnarray}
  \mmatrix[\normalsize]{q^E}_\nu&=&\big< \bm f_\nu^e,\bm E^{\rm inc}_2\big>\\
  \mmatrix[\normalsize]{q^H}_\nu&=&\big< \bm f_\nu^e,\bm H^{\rm inc}_2\big>\,,
\end{eqnarray}
\end{subequations}
and combine the electric and magnetic components in a single vector $q=(\mmatrix[\normalsize]{q_E},\mmatrix[\normalsize]{q_H})^T$, in complete analogy to Eq.~\eqref{eq:uvec}.  Together with the block matrices
\begin{equation}
  I=-\begin{pmatrix} \mmatrix[\normalsize]{I} & 0 \\ 0 & \mmatrix[\normalsize]{I} \end{pmatrix}\,,\quad
  A_j=\begin{pmatrix} \mmatrix[\normalsize]{\mathbb{D}_j} & -i\mu_0\omega \mmatrix[\normalsize]{\mathbb{S}_j} \\
  i\varepsilon_j\omega\mmatrix[\normalsize]{\mathbb{S}_j} & \mmatrix[\normalsize]{\mathbb{D}_j} \\ \end{pmatrix}\,,
\end{equation}
we are then led to our final Eq.~\eqref{eq:cal}.  For evaluating the boundary conditions within the Galerkin scheme, we introduce the matrices
\begin{eqnarray}
  \mmatrix{J}_{\nu\nu'} &=& \big< \bm f_\nu^e,\bm f_{\nu'}^e\big>\label{eq:matJ} \\
  \mmatrix{K}_{\nu\nu'} &=& \big<\nabla_\|\cdot\bm f_\nu^e,
  \nabla_\|\cdot\bm f_{\nu'}^e\big> \label{eq:matK}\,.
\end{eqnarray}
Inspection of Eq.~\eqref{eq:bctest} shows that we have to deal with a term of the form $\big<\nabla_\|\cdot\hat{\bm n}\times\bm f_\nu^e,\nabla_\|\cdot\bm f_{\nu'}^e\big>$ rather than with the expression of Eq.~\eqref{eq:matK}.  Any pairing bewteen two functions $\bm u$, $\bm v$ can be rewritten by inserting the identity operator expanded in the non-orthogonal basis of the shape elements via
\begin{equation}
  \big<\bm u,\bm v\big>=\sum_{\nu\nu'}\big<\bm u\,,\bm f_\nu^e\big>\left(\mmatrix[\normalsize]{J}^{-1}\right)_{\nu\nu'}
  \big<\bm f_{\nu'}^e,\bm v\big>\,.
\end{equation}
With this we then find $\big<\nabla_\|\cdot\hat{\bm n}\times\bm f_\nu^e,\nabla_\|\cdot\bm f_{\nu'}^e\big>=-\left(\mmatrix{I}\mmatrix{J}^{-1}\mmatrix{K}\right)_{\nu\nu'}$, where the negative sign is because of the reversed order of pairing functions in $\big<\hat{\bm n}\times\bm f_\nu^e,\bm f_{\nu'}^e\big>$ in comparison to the matrix of Eq.~\eqref{eq:matI}.  We are thus finally led to Eq.~\eqref{eq:bcworking} with the matrix
\begin{equation}\label{eq:matB}
  B_j=\begin{pmatrix} \mmatrix{J} & -\frac{i}{\omega\varepsilon_j}d_\perp\mmatrix[\normalsize]{\tilde K} \\
  i\omega\varepsilon_jd_\|\mmatrix{I} & \mmatrix{J} \\ \end{pmatrix}\,,
\end{equation}
where we have introduced $\mmatrix[\normalsize]{\tilde K}=-\mmatrix{I}\mmatrix{J}^{-1}\mmatrix{K}$.    

\end{appendix}


\begin{thebibliography}{42}
\bibitem[{com()}]{comment.cim}
\bibinfo{note}{In comparison to the contour integral method discussed in
  Ref.~\cite{hohenester:21} we have to compute the left eigenvectors
  numerically because the transmission matrix of Eq.~\eqref{eq:working} is no
  longer symmetric.}
\end{thebibliography}

\end{document}